\documentclass[aps,prb,twocolumn,superscriptaddress,showpacs,nofootinbib]{revtex4-2}
\usepackage[breaklinks,colorlinks,bookmarks=false,citecolor=blue,linkcolor=red,urlcolor=blue]{hyperref}
\usepackage{blindtext,enumitem,graphicx,dcolumn,color,xcolor,amsmath,braket,bm,changes,chemformula,cleveref,times}
\usepackage{mathtools}
\usepackage{booktabs}
\usepackage{multirow}
\usepackage{amssymb}
\usepackage{cleveref}
\graphicspath{{figures/}}
\usepackage{blindtext}
\usepackage{changes}
\usepackage{graphicx}
\usepackage[utf8]{inputenc}
\DeclareUnicodeCharacter{2032}{\ensuremath{\prime}}
\usepackage[english]{fixme}
\usepackage{xcolor}
\DeclareMathAlphabet{\mathbit}{OT1}{cmr}{bx}{it}
\usepackage{wasysym}            
\usepackage{tikz}
\usepackage[caption=false]{subfig}
\crefname{equation}{Eq.}{Eqs.}
\Crefname{equation}{Eq.}{Eqs.}
\crefname{figure}{Fig.}{Figs.}
\Crefname{figure}{Fig.}{Figs.}
\crefname{section}{Sec.}{Secs.}
\Crefname{section}{Sec.}{Secs.}

\begin{document}
\title{Detecting Symmetry-Resolved Entanglement: A Quantum Monte Carlo Approach}

\author{Kuangjie Chen}
		\affiliation{State Key Laboratory of Surface Physics, Institute of Nanoelectronics and Quantum Computing, Department of Physics, Fudan University, Shanghai 200438, China}
            \affiliation{Shanghai Qizhi Institute and Shanghai Artificial Intelligence Laboratory, Xuhui District, Shanghai 200232, China}


\author{Weizhen Jia}
		\affiliation{Department of Physics, The Chinese University of Hong Kong, Sha Tin, Hong Kong SAR, China}

\author{Xiaopeng Li}
\email{xiaopeng_li@fudan.edu.cn}
   \affiliation{State Key Laboratory of Surface Physics, Institute of Nanoelectronics and Quantum Computing, Department of Physics, Fudan University, Shanghai 200438, China}
\affiliation{Shanghai Qizhi Institute and Shanghai Artificial Intelligence Laboratory, Xuhui District, Shanghai 200232, China}
\affiliation{Shanghai Research Center for Quantum Sciences, Shanghai 201315, China} 
\affiliation{Hefei National Laboratory, Hefei 230088, China}

\author{René Meyer}
\email{rene.meyer@uni-wuerzburg.de}
		\affiliation{Institute for Theoretical Physics and Astrophysics and\\
Würzburg-Dresden Cluster of Excellence ct.qmat,\\
Julius-Maximilians-Universität Würzburg, 97074 Würzburg, Germany}
\affiliation{Shanghai Institute for Mathematics and Interdisciplinary Sciences (SIMIS), Shanghai, 200433, China}
        
\author{Jiarui Zhao}
        \email{jiaruizhao@cuhk.edu.hk}
		\affiliation{Department of Physics, The Chinese University of Hong Kong, Sha Tin, Hong Kong SAR, China}

\begin{abstract}
Symmetry and entanglement are two fundamental concepts in quantum many-body physics. Their interplay is captured by symmetry-resolved entanglement, which decomposes the total entanglement into contributions from different symmetry sectors. Computing symmetry-resolved entanglement in strongly interacting higher-dimensional quantum systems remains challenging. Here, we formulate and implement an estimator-based quantum Monte Carlo (QMC) framework for computing symmetry-resolved R\'enyi entropies (SRRE) in sign-problem-free interacting lattice systems by measuring disorder (symmetry-twisted) operators in ordinary and replica ensembles and reconstructing SRRE from the corresponding charged moments. We validate the framework in two controlled one-dimensional settings: the transverse-field Ising model (TFIM), for which exact conformal-field-theory predictions are available, and the interacting Heisenberg chain, which tests the $U(1)$ symmetry-sector reconstruction and its finite-size behavior. We then apply the method to the two-dimensional TFIM. Within the accessible system sizes and a phenomenological finite-size extrapolation, our data provide numerical evidence consistent with entanglement equipartition at the $(2+1)$D Ising critical point. Our work establishes a practical numerical route to symmetry-resolved entanglement in interacting lattice models and provides a framework for future studies beyond one dimension.
\end{abstract}

\maketitle

\noindent{\textcolor{blue}{\it Introduction.}---} 
Entanglement entropy (EE) is a powerful probe of quantum many-body states, offering insights that transcend the paradigm of conventional order parameters~\cite{Pasquale_Calabrese_2004, RevModPhys.80.517, Laflorencie_2016}. Its scaling with subsystem size offers an effective organizing principle, uncovering universal features of distinct quantum phases. While EE typically obeys an area law in ground states of local gapped Hamiltonians~\cite{RevModPhys.82.277}, deviations from this behavior often signal more exotic forms of quantum matter. For example, topological entanglement entropy characterizes topologically ordered phases~\cite{kitaevTopological2006,IsakovTopological2011}, while logarithmic corrections carrying universal information can arise in (1+1)D conformal field theories (CFTs), in systems with Nambu-Goldstone modes, and for entangling boundaries with corners~\cite{kulchytskyyDetecting2015,zhaoMeasuring2022,zhaoScaling2022,song2023deconfined,EmidioUniversal2024}. As a result, entanglement scaling has become a valuable framework not only for classifying equilibrium phases, but also for diagnosing nonequilibrium phenomena such as measurement-induced phase transitions, which emerge from the interplay between unitary evolution and local measurements~\cite{SkinnerPRX2019,ROeland2023,Paviglianiti2023ee}.

In recent years, this framework has been extended by incorporating the role of global symmetries, yielding the notion of \textit{symmetry-resolved entanglement}~\cite{Laflorencie_2014,PhysRevLett.120.200602,PhysRevB.98.041106,Bonsignori_2019,Capizzi_2020,Estienne_2021,digiulio2023boundaryconformalfieldtheory,bueno2022universal,huang2025symmetryresolvedentanglemententropyhigher}. A central result is the \textit{equipartition of entanglement}~\cite{PhysRevB.98.041106,digiulio2023boundaryconformalfieldtheory}: in the settings where it holds, the leading asymptotic contribution to the symmetry-resolved entropy becomes independent of the sector label, although the sector probabilities and subleading corrections may retain symmetry dependence. Symmetry-resolved entanglement therefore provides a refined probe of the entanglement structure and universal behavior of quantum states. Current research frontiers are expanding its scope to systems with non-Abelian symmetries~\cite{10.21468/SciPostPhys.17.5.127}, nonequilibrium dynamics~\cite{Parez2021Exact}, multipartite entanglement~\cite{jain2025}, and measurements in quantum simulators~\cite{Lukin_2019}.

Symmetry-resolved entanglement in one dimension has already been studied extensively using conformal field theory methods, exact free-fermion techniques, density matrix renormalization group (DMRG), and related tensor-network approaches~\cite{Laflorencie_2014,PhysRevB.98.041106,Bonsignori_2019,FraenkelGoldstein2020,SCHOLLWOCK201196}. Accordingly, the one-dimensional TFIM and Heisenberg chain considered below are not intended as systems inaccessible to established methods. Instead, they provide controlled benchmarks of two complementary components of our formulation: the $\mathbb{Z}_2$ charged-moment estimator on a replica manifold against an exact CFT prediction, and the $U(1)$ Fourier reconstruction in an interacting model.


Beyond one dimension, symmetry-resolved entanglement has been analyzed in several settings with additional analytical structure, including higher-dimensional free Fermi gases treated by exact or multidimensional-bosonization methods, two-dimensional free fermions and bosons treated by dimensional reduction, and higher-dimensional $U(1)$-symmetric conformal field theories~\cite{FraenkelGoldstein2020,TanRyu2020,Murciano_2020,huang2025symmetryresolvedentanglemententropyhigher}. These studies provide important analytical expectations and establish equipartition in their respective regimes. Complementary work has connected symmetry-resolved entanglement to symmetry-protected and topological crystalline phases, developing correlation-function methods and deriving lower bounds on the number and configurational entropies from topological invariants~\cite{MonkmanSirker2023General,PhysRevB.107.125108}. However, these analytically structured settings do not directly address the on-site Ising $\mathbb{Z}_2$ symmetry and cylindrical bipartition studied here through replica-manifold charged moments, nor do they provide a general scalable numerical route for interacting lattice critical points. Direct numerical access to replica symmetry-resolved quantities in this latter setting remains limited.

Subsystem disorder operators have already been developed as quantum Monte Carlo (QMC) probes of quantum criticality in a sequence of earlier works. For the $(2+1)$D Ising transition, Ref.~\cite{zhaoHigher2021} evaluated the same class of $\mathbb{Z}_2$ symmetry-twist operator in the ordinary QMC ensemble and established perimeter-law scaling with multiplicative finite-size corrections. This program was subsequently extended to $U(1)$-symmetric conventional critical points and deconfined quantum criticality, where disorder-operator scaling and corner corrections were used to extract universal current-response information~\cite{wangScaling2021,wangScaling2022}. A recent iPEPS study provides a complementary finite-correlation-length analysis of the ordinary single-replica disorder parameter~\cite{XuHuang2025}.

In this article, we formulate and implement an estimator-based QMC framework for computing symmetry-resolved R\'enyi entropies in sign-problem-free interacting systems beyond one dimension. Building on the charged-moments framework~\cite{PhysRevLett.120.200602}, we relate charged moments to subsystem disorder operators measured in ordinary and glued replica ensembles and reconstruct the symmetry-resolved moments by Fourier transformation. The essential extension relative to the earlier disorder-operator QMC program is the measurement on replica manifolds, where the symmetry-twisted observables directly provide the charged moments required for symmetry resolution. A practical advantage is that one simulation in the ordinary ensemble and one in the replica ensemble provide all required counting fields and multiple symmetry sectors, rather than requiring an independent simulation for every counting field or charge sector. The one-dimensional systems serve as controlled validations of this procedure, while the two-dimensional TFIM illustrates its main intended application in an interacting system beyond one dimension.

The remainder of the paper is organized as follows. We first introduce symmetry-resolved entanglement and entanglement equipartition. We then present the general QMC framework and the models studied in this work. After discussing the one- and two-dimensional numerical results, we summarize the limitations and future directions. In Appendix~\ref{app:Benchmark}, we benchmark the QMC implementation against exact diagonalization. Appendix~\ref{app:U1Reconstruction} documents the replica implementation, the $U(1)$ symmetry-sector reconstruction, stability checks, and the remaining numerical details required for reproducibility.



\noindent{\textcolor{blue}{\it Symmetry-Resolved Entanglement Entropy.}---} 

\begin{figure}[htp]
\centering
\includegraphics[width=\columnwidth]{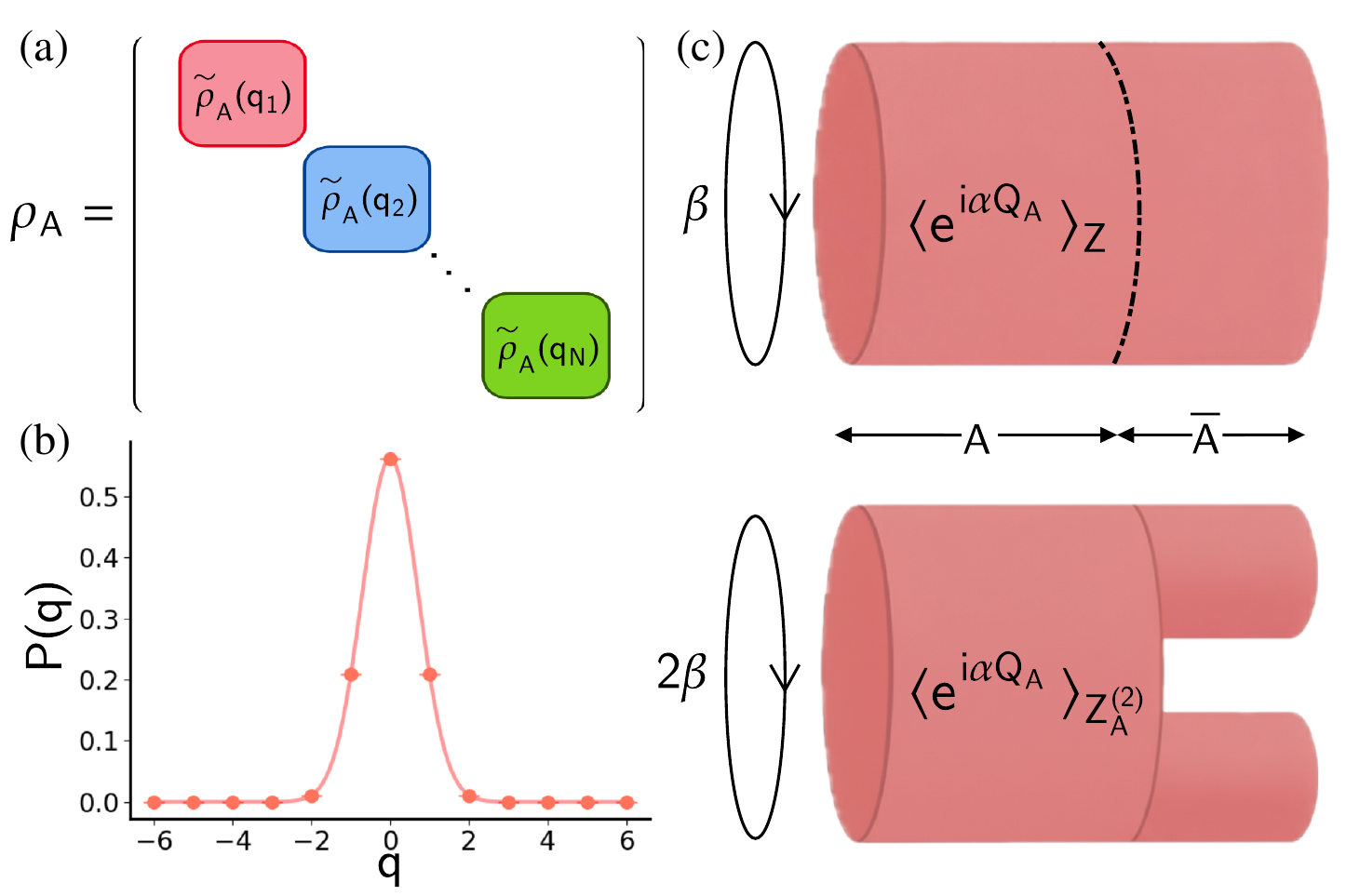}
\caption{(a) Block-diagonal structure of $\rho_A$ in the eigenbasis of the conserved charge $Q_A$. (b) Typical probability distribution $P(q)=\mathrm{Tr}[\tilde{\rho}_A(q)]$ in the 1D Heisenberg model, with $q=S_A^z$. (c) Single-replica (upper panel) and two-replica (lower panel) configurations corresponding to the ordinary partition function $Z$ and the replica partition function $Z_A^{(2)}$, respectively. The geometry of $Z_A^{(2)}$ can be viewed as two replicas of the single-replica configuration, with region $A$ glued together along the imaginary-time direction. The disorder operator $e^{i\alpha Q_A}$ is measured in the entangling region $A$.}
\label{fig:fig1}
\end{figure}

\noindent
For a bipartite quantum system described by the density matrix $\rho$, the entanglement between subsystem $A$ and its complement $\bar{A}$ is encoded in the reduced density matrix (RDM) $\rho_A=\mathrm{Tr}_{\bar{A}}\rho $. The von Neumann entropy $ S_{A}^{(vN)} = -\mathrm{Tr}(\rho_A \ln \rho_A) $ is a well-established entanglement measure, and its generalization, the $n$-th Rényi entanglement entropy $ S_A^{(n)} = \frac{1}{1-n} \ln \mathrm{Tr}[\rho_A^n] $, reduces to the von Neumann entropy in the limit $ n \to 1 $. If the system possesses a global symmetry with a conserved charge $ Q = Q_{A} + Q_{\bar{A}}$ where $Q$ is a sum of local operators, it is then straightforward to show that $ [\rho_A, Q_A] = 0 $, which implies that $\rho_A $ can be block-diagonalized with respect to the eigenspaces of $Q_A$. As shown in Fig.~\ref{fig:fig1}(a), $\rho_A$ can be expressed as
\begin{equation}
\rho_A=\bigoplus_q \tilde{\rho}_{A}(q)\,,
\end{equation}
where $\tilde{\rho}_{A}(q)$ is the RDM block labeled by eigenvalue $q$. We can also rewrite $\rho_A$ as 
\begin{equation}
\rho_A = \bigoplus_q  P(q) \rho_A(q)\,,
\end{equation}
where $\rho_A(q)=\tilde{\rho}_A(q)/\mathrm{Tr} [\tilde{\rho}_A(q)]$ is the normalized RDM block and $P(q)=\mathrm{Tr}[\tilde{\rho}_A(q)]$ can be interpreted as the probability of finding eigenvalue $q$ in a measurement of $Q_{A}$. An example of  $P(q)$ for 1D Heisenberg chain with $L=768$ that will be discussed later is shown in Fig.~\ref{fig:fig1} (b).

In analogy with the definition of R\'enyi EE $S_A^{(n)}$, the symmetry-resolved R\'enyi entropy (SRRE) in each charge sector is defined as~\footnote{In the following, we omit the subscript $A$ for simplicity.}
\begin{equation}
S_n(q) = \frac{1}{1-n}\ln \mathrm{Tr}\!\left[\rho_A(q)^{n}\right]\,,
\end{equation}
which also converges to the symmetry-resolved von Neumann entropy as $n\rightarrow 1$:
\begin{equation}
\lim_{n \to 1}S_{n}(q) =S_{vN}(q)\coloneqq -\mathrm{Tr} [\rho_A (q) \ln \rho_A(q)]\,.
\end{equation} 
Note that for von Neumann entropy, the following decomposition relation holds:
\begin{equation}
    S_{A}^{(vN)} =  \sum_{q} P(q) S_{vN}(q) - \sum_{q} P(q) \ln P(q)\,,
\end{equation}
where the first term captures the contributions from individual charge sectors and the second term arises from charge fluctuations~\cite{Lukin_2019, PhysRevResearch.2.043191}. 
As discussed in \cite{digiulio2023boundaryconformalfieldtheory}, such a decomposition only holds for the Rényi entropy if equipartition of entanglement holds in the system under consideration.  

Physically, equipartition does not in general require the probabilities $P(q)$ of different sectors to be equal. Rather, after the reduced density matrix is normalized within each sector, the leading asymptotic contributions to $S_n(q)$ become independent of the sector label; sector dependence may remain in $P(q)$ and in subleading finite-size corrections. Equipartition is therefore a sharper universal property than the numerical coincidence of sector-resolved curves and makes symmetry-resolved entanglement a refined probe of critical behavior. Related sector-resolved information is also being explored for non-Abelian symmetries, nonequilibrium dynamics, multipartite entanglement, and quantum-simulator measurements~\cite{10.21468/SciPostPhys.17.5.127,Parez2021Exact,jain2025,Lukin_2019}.

Symmetry-resolved entanglement has been considered in one spatial dimension for critical spin chains and Luttinger liquids in \cite{Laflorencie_2014} and for free fermions in \cite{Bonsignori_2019}. Equipartition of entanglement is the observation that in these systems, at least\footnote{In WZW models, equipartition is broken to constant order \cite{calabrese2021symmetry}, and for the $c=1$ compact boson, equipartition was found to hold to all orders in the UV cutoff expansion upon imposing appropriate (NN and DD) boundary conditions in the BCFT framework in \cite{digiulio2023boundaryconformalfieldtheory}.} the first two terms in the UV cutoff expansion of the symmetry-resolved entanglement entropy are independent of the subregion charge. Instead, they are determined by the central charge $c$ and the Kac-Moody level $k$ of the underlying $U(1)_k$ symmetry algebra alone \cite{PhysRevB.98.041106},
\begin{equation}\label{eq:equipart1}
    S_{vN}(q) = \frac{c}{3}\ln \frac{\ell_A}{\epsilon} - \frac{1}{2}\ln \left(\frac{k}{\pi} \ln \frac{\ell_A}{\epsilon}\right) + {\cal O}(1)\,.
\end{equation}
For $U(1)_k$, the constant contribution in ${\cal O}(1)$ is also charge-independent, while for WZW models with symmetry group $G$ it depends on the dimension of the group representation chosen \cite{calabrese2021symmetry}. Equipartition up to constant order can be traced back to the structure of the $U(1)_k$ Kac-Moody algebra \cite{zhao2021symmetry}, and also holds in excited states \cite{weisenberger2021symmetry}. The universality up to double-logarithmic order is tied to the twist operators \cite{Pasquale_Calabrese_2004} acquiring a universal shift in dimension after inserting the charged defect operator $e^{i\alpha Q_A}$, which only depends on the structure of the symmetry algebra and the ensuing Fourier transformation to the charged partition function. On the other hand, equipartition was found to be broken in the leading order logarithmic term already by analysing the charged moments \eqref{eq:charged-moments} at small chemical potentials $\alpha$ for 2D CFTs with $W_3$ higher spin symmetry in \cite{zhao2022charged}. The  works \cite{zhao2021symmetry,weisenberger2021symmetry} also established a holographic prescription to calculate symmetry-resolved entanglement entropy in terms of a bulk gauge field Wilson line \cite{belin2013holographic} via the charged moment \eqref{eq:charged-moments} defined below. Higher order terms in the UV cutoff expansion come with positive powers of $\frac{\epsilon}{\ell_A}$ and are, as is the ${\cal O}(1)$ contribution, non-universal, i.e. cutoff dependent. Nevertheless, applying highly symmetric boundary conditions such as the conformal boundary conditions $T=\bar T$ and Neumann or Dirichlet-type conditions $J = \pm \bar J$, respectively, equipartition can be shown to hold to all orders in the UV cutoff expansion for $U(1)_k$ Kac-Moody algebra  \cite{digiulio2023boundaryconformalfieldtheory}.  

\noindent{\textcolor{blue}{\it General Methods.}---} 
Our method starts with the charged moment defined as~\cite{PhysRevLett.120.200602}
\begin{equation}
\label{eq:charged-moments}
Z_n(\alpha) =  \mathrm{Tr} (\rho_A^{\,n}\,e^{i\alpha Q_A})\,.
\end{equation}
 Treating $\alpha$ as a continuous variable in $(-\pi,\pi)$, the Fourier transform of $Z_n(\alpha)$ yields
\begin{equation}
\label{eq:fouier-transformation}
Z_n(q) = \int_{-\pi}^{\pi}\frac{d\alpha}{2\pi}e^{-i\alpha q}Z_n(\alpha)
=\mathrm{Tr}(\Pi_q\rho_A^n)\,,
\end{equation}
where $\Pi_q$ is the projector onto the subspace with $Q_{A}=q$.  

For a discrete $\mathbb{Z}_N$ symmetry, the integral in Eq.~\eqref{eq:fouier-transformation} is replaced with a discrete Fourier transform 
\begin{equation}
\label{eq:sector-moment-ZN}
Z_n(q)=\frac{1}{N}\sum_{k=0}^{N-1} e^{-i\frac{2\pi k}{N}q}\;
Z_n\!\left(\alpha=\frac{2\pi k}{N}\right)\
=\mathrm{Tr}(\Pi_q\rho_A^n),\
\end{equation}
where $q=0,...,N-1$ and $\alpha$  can only take discrete values $2\pi k/N$ with $k=0,...,N-1$. 

It then follows immediately that 
\begin{equation}
Z_1(q)=\mathrm{Tr}(\Pi_q\rho_A) \equiv P(q),
\end{equation}
and
\begin{equation}
Z_n(q)=P(q)^n\,\mathrm{Tr}\!\left[\rho_A(q)^n\right].
\end{equation}
Thus the $n$-th SRRE reads
\begin{equation}
\label{eq:SREE-definition}
S_n(q) = \frac{1}{1-n}\ln\!\frac{Z_n(q)}{\big[Z_1(q)\big]^n}\,.
\end{equation}

Based on the above equations, the \textit{core} of our method is to measure $Z_1(\alpha)$ and $Z_n(\alpha)$ in the ordinary and replica QMC ensembles and then obtain $Z_1(q)$ and $Z_n(q)$ by Fourier transforming the charged moments according to Eq.~\eqref{eq:fouier-transformation}. A practical advantage is that $e^{i\alpha Q_A}$ is a physical observable on the $n$-sheeted replica manifold. The complete subtracted SRRE therefore requires one simulation in the ordinary ensemble and one in the corresponding replica ensemble; within each ensemble, the observables for all required counting fields can be accumulated from the same sampled configurations and multiple symmetry sectors can be reconstructed without an independent simulation for every $\alpha$ or $q$.

For $n=1$, the charged moment reduces to $Z_1(\alpha)=\mathrm{Tr}(\rho_A e^{i\alpha Q_A})=\langle e^{i\alpha Q_A}\rangle_Z$, namely the ordinary subsystem disorder operator illustrated in Fig.~\ref{fig:fig1}(c). Such observables were previously studied using QMC at $\mathbb{Z}_2$, $U(1)$, and deconfined quantum critical points~\cite{zhaoHigher2021,wangScaling2021,wangScaling2022}. The essential extension introduced here is their measurement for $n\geq2$ on glued replica manifolds, where they provide the normalized charged moments required for symmetry resolution.

For $n\ge 2$, we have
\begin{equation}
\label{Zn_alpha}
Z_n(\alpha)=\frac{\mathrm{Tr}(\rho_A^n e^{i\alpha Q_A})}{\mathrm{Tr}(\rho_A^n)}\,\mathrm{Tr}(\rho_A^n)
=\langle e^{i\alpha Q_A} \rangle_{Z_A^{(n)}} e^{(1-n)S_A^{(n)}}\,,
\end{equation}
where $\langle \cdots \rangle_{Z_A^{(n)}}$ denotes the normalized expectation value in the QMC ensemble defined on the corresponding $n$-sheeted replica manifold. Equivalently, if $Z$ is the ordinary partition function and $Z_A^{(n)}$ denotes the partition function on the glued replica manifold, then $\mathrm{Tr}(\rho_A^n)=Z_A^{(n)}/Z^n$. In this work we focus on the case $n=2$. The corresponding simulation is performed in a two-replica ensemble, in which subsystem $A$ is glued together along the imaginary-time direction, as illustrated in Fig.~\ref{fig:fig1}(c).

Numerically, we employ the stochastic series expansion in a basis where the relevant subsystem symmetry operator is diagonal. The TFIM is simulated in the $\sigma^x$ basis using directed-loop updates, while the Heisenberg chain is simulated in the $S^z$ basis using operator-loop updates~\cite{Sandvik_2002,Sandvik1999}. For each model, the ordinary and two-replica ensembles use the same model-specific update machinery. The replica geometry differs through the imaginary-time boundary connectivity: degrees of freedom in region $A$ are connected between the two replicas, whereas those in $\bar{A}$ close within each replica. The methodological ingredient specific to the present framework is therefore the charged-moment measurement on the replica manifold and the subsequent symmetry-sector reconstruction, rather than a new Monte Carlo update for each counting field or sector. In the $U(1)$ case, $Q_A$ is evaluated once on each sampled configuration and all $\cos(\alpha_jQ_A)$ are accumulated simultaneously. Further implementation and reconstruction details are given in Appendix~\ref{app:U1Reconstruction}.

\begin{figure*}[htp!]
  \centering
  \includegraphics[width=2\columnwidth]{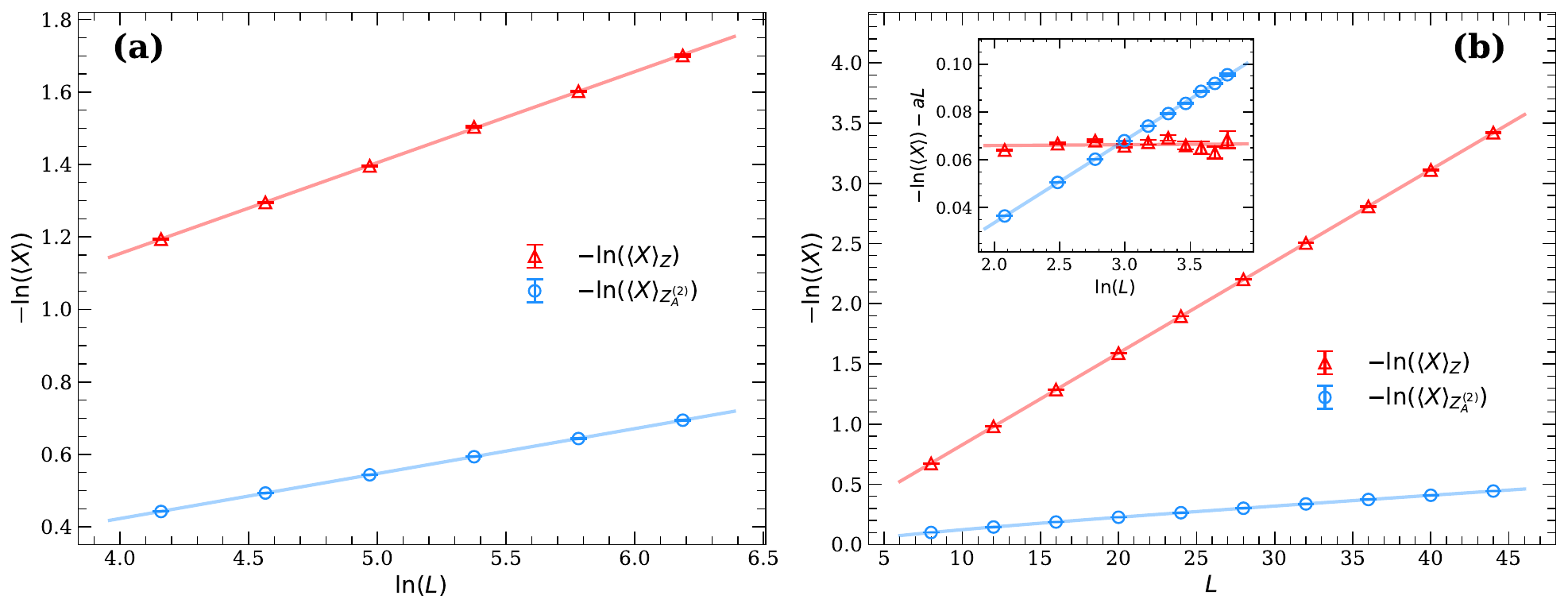}
  \caption{
Scaling of the disorder operator $\langle X \rangle$ in the single- and two-replica ensembles at the quantum critical point of the TFIM.
(a) 1D chain of length $L$ with a half-chain entangling region ($L_A=L/2$). The data are fitted by
$-\ln \langle X \rangle_Z = 0.251(1)\ln L + 0.149(5)$
and
$-\ln \langle X \rangle_{Z_A^{(2)}} = 0.1242(4)\ln L - 0.074(2)$.
(b) 2D $L\times L$ square lattice with an entangling region of size $L/2\times L$ and a smooth boundary. The single-replica data are fitted by
$-\ln \langle X \rangle_Z = 0.0762(1)L + 0.066(2)$,
while the two-replica data are described over the accessible sizes by the phenomenological form
$-\ln \langle X \rangle_{Z_A^{(2)}} = 0.00793(2)L + 0.0343(3)\ln L - 0.0347(5)$.
The inset shows $-\ln \langle X \rangle - aL$ versus $\ln L$ for both ensembles, where $a$ denotes the fitted coefficient of the leading linear term.
}
  \label{fig:fig2}
\end{figure*}

According to Eqs.~\eqref{eq:fouier-transformation}, \eqref{eq:SREE-definition}, and \eqref{Zn_alpha}, for continuous $\alpha$ we obtain
\begin{align}
\label{eq:subtractedRee}
S_n(q)-S_{A}^{(n)}={}&
\frac{n}{n-1}\ln \int_{-\pi}^{\pi}\frac{d\alpha}{2\pi}e^{-i\alpha q}\langle e^{i\alpha Q_A}\rangle_{Z}\\
&+\frac{1}{1-n} \ln \int_{-\pi}^{\pi}\frac{d\alpha}{2\pi}e^{-i\alpha q} \langle e^{i\alpha Q_A} \rangle_{Z_A^{(n)}}\,. \nonumber
\end{align}
For discrete $\alpha$, we simply replace the Fourier transform in its discrete form.

Since the scaling forms of $S_{A}^{(n)}$ for a wide variety of quantum systems are already well understood, our task is reduced to measuring $\langle e^{i\alpha Q_A}\rangle_{Z}$ and $\langle e^{i\alpha Q_A} \rangle_{Z_A^{(n)}}$ separately in two QMC simulations, from which the difference between the symmetry-resolved R\'enyi entropy and the total R\'enyi EE can be determined. We will denote this difference as \textit{subtracted} SRRE in the rest of the paper. 

When the exact values of $S_n(q)$ are needed, $S_{A}^{(n)}$ itself can also be computed independently using the recently developed nonequilibrium methods based on Jarzynski equality~\cite{Jarzynski1997} within the QMC framework, which have proven to be highly efficient in both boson and fermion systems~\cite{demidioEntanglement2020,zhaoMeasuring2022,PanGaopei2023, EmidioUniversal2024}. Our methods are therefore broadly applicable in quantum Monte Carlo simulations.

\noindent{\textcolor{blue}{\it Models.}---}
\noindent The first model we consider in this work is the transverse-field Ising model (TFIM) in 1D and 2D with the following Hamiltonian:
\begin{equation}
    H=-J\sum_{\langle i,j \rangle}\sigma_{i}^{z}\sigma_{j}^{z}-h\sum_{i}\sigma_{i}^{x}\,,
\end{equation}
where $\langle i,j \rangle$ refers to the nearest-neighbors and the Hamiltonian possesses the $\mathbb{Z}_2$ global symmetry. The system undergoes an ordered-to-disordered phase transition when tuning the magnitude of $h/J$. The $\mathbb{Z}_2$ symmetry is spontaneously broken inside the ordered phase and maintained both at the critical point and inside the disordered phase. We perform all calculations at the corresponding critical points of the model, namely $h_c/J=1$ for the 1D chain and $h_c/J\approx 3.044$ for the square lattice~\cite{Bloete2002, ZihongItinerant2019}.




For the TFIM, the global $\mathbb{Z}_2$ spin-flip symmetry splits the Hilbert space of subsystem $A$ into even ($q=0$) and odd ($q=1$) parity sectors. The corresponding symmetry operator restricted to $A$ is $X=\prod_{i\in A}\sigma_i^x$. Consequently, the discrete Fourier transform in Eq.~\eqref{eq:sector-moment-ZN} reduces to two charged moments only, evaluated at $\alpha=0$ and $\alpha=\pi$, corresponding to the identity operator and $X$, respectively.

From the two key observables, $\langle X \rangle_Z$ and $\langle X \rangle_{Z_A^{(2)}}$, we can reconstruct the symmetry-resolved moments for the charge sectors $q=0,1$:
\begin{subequations}\label{eq:ZQA_polished}
\begin{align}
  Z_1(q) &= \tfrac{1}{2}\bigl(1+(-1)^{q}\langle X \rangle_Z\bigr)\,, \label{eq:Z1_polished}\\
  Z_2(q) &= \tfrac{1}{2}\,e^{-S_A^{(2)}}\bigl(1+(-1)^{q}\langle X \rangle_{Z_A^{(2)}}\bigr)\,. \label{eq:Z2_polished}
\end{align}
\end{subequations}
Substituting these expressions into the Eq.~\eqref{eq:SREE-definition} yields the closed-form result:\
\begin{equation}
S_2(q)-S_A^{(2)}= -\ln\!\left[\frac{1+(-1)^{q}\langle X \rangle_{Z_A^{(2)}}}{\bigl(1+(-1)^{q}\langle X \rangle_Z\bigr)^2}\right]-\ln 2\,.
 \label{eq:subSREE}
\end{equation}

Therefore, for the $\mathbb{Z}_2$ TFIM, the subtracted symmetry-resolved R\'enyi entropy can be obtained directly from disorder operator expectation values measured in the single- and two-replica ensembles, without explicitly performing the Fourier transform or separately evaluating the total R\'enyi EE.

Equation~\eqref{eq:subSREE} also makes the asymptotic criterion for $\mathbb{Z}_2$ equipartition explicit. If the nontrivial charged moments $\langle X\rangle_Z$ and $\langle X\rangle_{Z_A^{(2)}}$ vanish in the thermodynamic limit, then $P(0)=P(1)\to1/2$ and
\[
S_2(q)=S_A^{(2)}-\ln 2+o(1),\qquad q=0,1.
\]
Thus, the even- and odd-parity sectors inherit the same leading second-R\'enyi entanglement scaling, while their difference is confined to finite-size corrections controlled by the symmetry-twisted moments. Equivalently, equipartition corresponds to the asymptotic suppression of the nontrivial $\mathbb{Z}_2$-twisted contribution relative to the identity contribution in the symmetry projection.

The second model we consider is the spin-$\frac{1}{2}$ antiferromagnetic Heisenberg chain, whose low-energy physics is described by a $(1+1)$D conformal field theory. The Hamiltonian reads
\begin{equation}
    H=J\sum_{i=1}^{N} \Vec{S}_{i}\cdot \Vec{S}_{i+1}\,,
\end{equation}
with periodic boundary conditions, $\Vec{S}_{N+1}=\Vec{S}_{1}$. Although the model exhibits full $SU(2)$ spin-rotational symmetry, in this work we focus on the $U(1)$ subgroup corresponding to spin rotations about the $z$ axis. The corresponding symmetry charge in subsystem $A$ is the total magnetization,$Q_A=S_A^z=\sum_{i\in A}S_i^z$, whose eigenvalues range from $-L_A/2$ to $L_A/2$. Because the subsystem charge takes discrete values on a finite lattice, the symmetry-resolved moments, and hence the subtracted SRRE, are reconstructed from the charged moments through a discrete Fourier inversion.

\begin{figure*}[htp!]
\centering
\includegraphics[width=2\columnwidth]{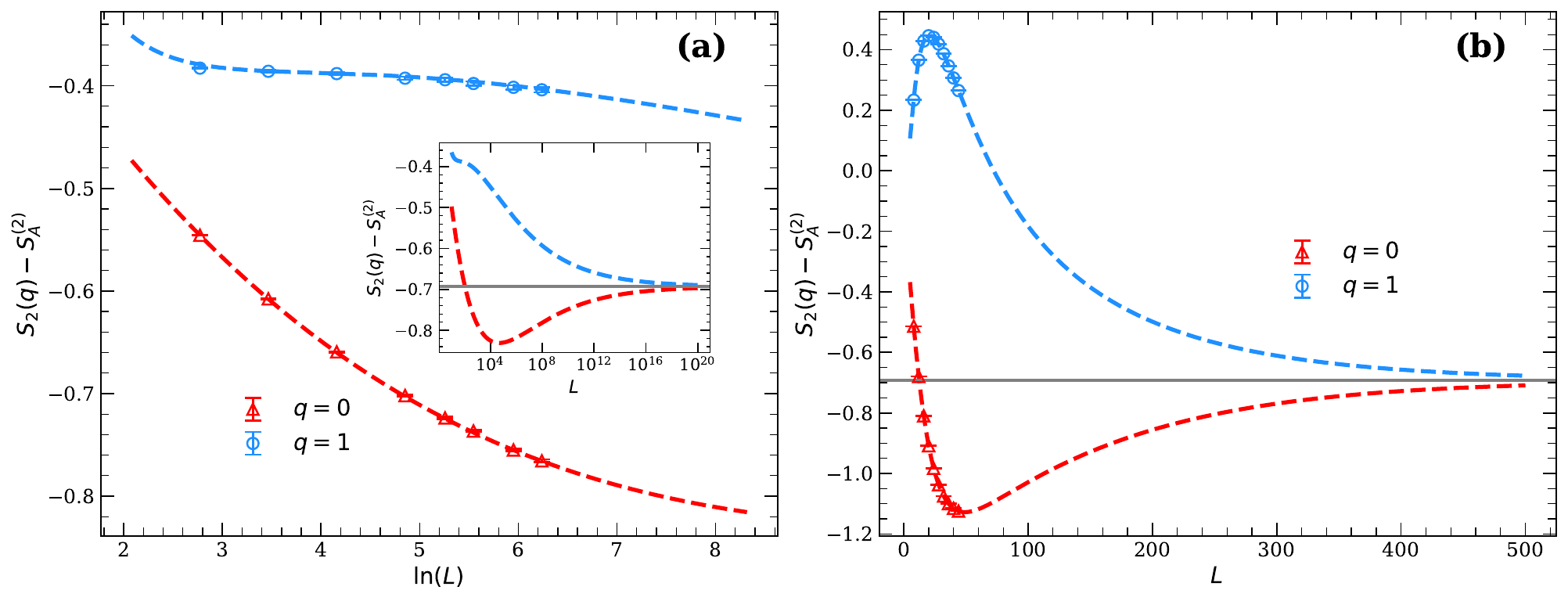}
\caption{Asymptotic scaling of the subtracted symmetry-resolved R\'enyi entropy $S_2(q)-S_A^{(2)}$ in the 1D (panel (a)) and 2D (panel (b)) TFIM at their respective QCPs. Symbols denote directly simulated numerical data. The dashed curves are extrapolations obtained by inserting the fitted scaling forms from Fig.~\ref{fig:fig2} into Eq.~\eqref{eq:subSREE}; they do not represent additional directly simulated system sizes. The inset in panel (a) shows the extrapolated large-$L$ behavior.}
\label{fig:fig3}
\end{figure*}

\noindent{\textcolor{blue}{\it Results.}---} 

For the models introduced above, we use stochastic series expansion (SSE) QMC simulations in the ordinary and two-replica ensembles. The TFIM is simulated in the $\sigma^x$ basis using directed-loop updates, while the Heisenberg chain is simulated in the $S^z$ basis using operator-loop updates~\cite{Sandvik_2002,Sandvik1999}. The same model-specific updates are used in the corresponding replica ensembles, with the imaginary-time boundary connectivity modified according to the glued geometry of region $A$. In 1D, we consider periodic chains of length $L$ and choose a half-chain entangling region, $L_A=L/2$. In 2D, we study $L\times L$ square lattices with periodic boundary conditions and take $A$ to be a cylindrical subregion of size $L/2\times L$ with a smooth (corner-free) boundary.

We first benchmark the $\mathbb{Z}_2$ replica-manifold estimator on the 1D TFIM at the $(1+1)$D Ising quantum critical point ($h_c/J=1$), using $\beta=4L$ to access ground-state properties. In the $(1+1)$D Ising CFT, the disorder operator $X=\prod_{i\in A}\sigma_i^x$ scales as

\begin{equation}
-\ln\langle X \rangle_n = \frac{2(\Delta_{\nu}+\bar{\Delta}_{\nu})}{n} \ln L = \frac{1}{4n}\ln L,
\end{equation}
where $\Delta_{\nu}=\bar{\Delta}_{\nu}=1/16$ are the scaling dimensions of the disorder operator, and $n$ denotes the R\'enyi index of the replica ensemble~\cite{PhysRevLett.120.200602}. As shown in Fig.~\ref{fig:fig2}(a), the extracted logarithmic prefactors are $0.251(1)$ for the single-replica ensemble ($n=1$) and $0.1242(4)$ for the two-replica ensemble ($n=2$). These values are in excellent agreement with the exact predictions, $1/4=0.25$ and $1/8=0.125$, respectively. The small deviation of the $n=2$ result from the exact value is likely attributable to finite-size effects.



We next study the 2D TFIM at the $(2+1)$D Ising quantum critical point ($h_c/J\approx3.044$), using $\beta=2L$. For an extended disorder operator in two spatial dimensions, the leading contribution is expected to obey a perimeter law, $\langle X\rangle\sim C e^{-aL}$. As shown in Fig.~\ref{fig:fig2}(b), the single-replica data recover the perimeter-law scaling previously established by QMC for the $(2+1)$D Ising critical point~\cite{zhaoHigher2021}; a recent iPEPS finite-correlation-length study provides a complementary tensor-network characterization of the same ordinary disorder parameter~\cite{XuHuang2025}. Our fitted single-replica coefficient is $0.0762(1)$. By contrast, the two-replica data display a systematic deviation from a pure perimeter law. After subtracting the leading linear contribution, the residual is approximately linear in $\ln L$ over the accessible system sizes, as illustrated in the inset of Fig.~\ref{fig:fig2}(b).

Motivated by this numerical observation and by multiplicative power-law corrections found in related studies of disorder operators for rectangular subregions~\cite{zhaoHigher2021,wangScaling2021,wangScaling2022}, we consider the phenomenological form
\[
\langle X\rangle_{Z_A^{(2)}}\sim C e^{-aL}L^{-b},
\]
or equivalently
\begin{equation}
-\ln \langle X \rangle_{Z_A^{(2)}} = aL + b\ln L + c\,.
\label{eq:2Dfit}
\end{equation}
We emphasize that Eq.~\eqref{eq:2Dfit} is a phenomenological parametrization of the accessible size range rather than a first-principles derivation of the asymptotic scaling for the present smooth-boundary replica geometry. In the related rectangular geometries, logarithmic terms are commonly associated with geometric singularities such as corners, whereas the entangling boundary considered here is smooth. Other forms of subleading finite-size corrections therefore cannot presently be excluded. Determining the asymptotic form and the possible universality of the apparent logarithmic contribution requires larger-scale simulations and further theoretical analysis, which we leave for future work~\cite{Chen2026Renyi}.

Substituting the fitted scaling functions of $\langle X\rangle_Z$ and $\langle X\rangle_{Z_A^{(2)}}$ into Eq.~\eqref{eq:subSREE} gives the dashed curves in Fig.~\ref{fig:fig3}. The symbols in that figure are directly simulated data, whereas the dashed curves represent extrapolations. For the 1D TFIM, the CFT-consistent extrapolation indicates that system sizes of order $L\sim10^{20}$ would be required for clear convergence to $-\ln2$. For the 2D TFIM, the extrapolation based on the phenomenological form in Eq.~\eqref{eq:2Dfit} suggests an approach to the asymptotic value on length scales of order $L\sim500$, substantially faster than in 1D; this scale is not directly simulated. The qualitative $\mathbb{Z}_2$ equipartition criterion following Eq.~\eqref{eq:subSREE} does not require the logarithmic term to be the unique asymptotic correction: Eq.~\eqref{eq:2Dfit} is used to characterize the finite-size approach and to provide an indicative extrapolation scale. The directly simulated data and the corresponding finite-size analysis therefore provide numerical evidence consistent with second-R\'enyi entanglement equipartition at the Ising quantum critical points.

\begin{figure}[htp!]
\centering
\includegraphics[width=1\columnwidth]{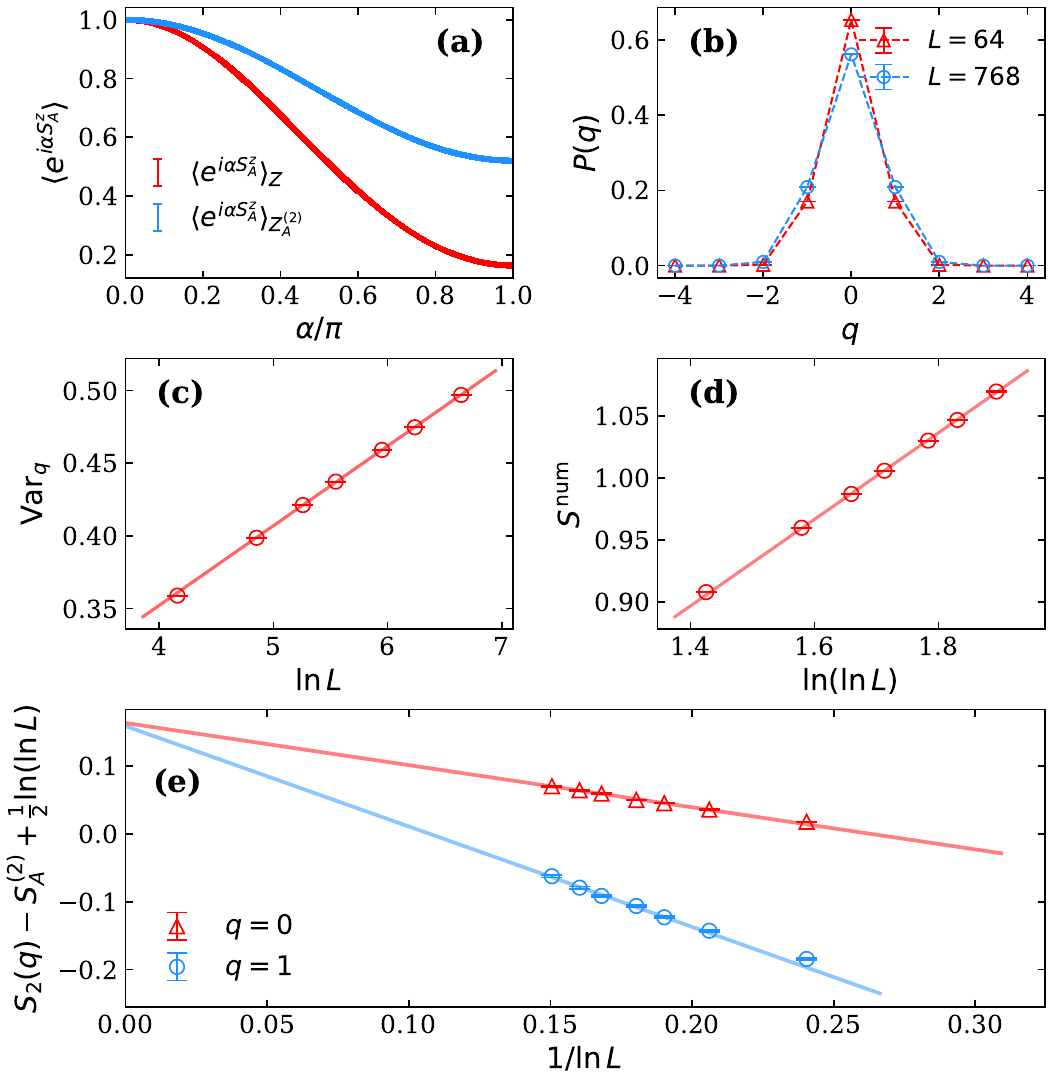}
\caption{
Analysis of symmetry-resolved entanglement for the 1D Heisenberg model. 
(a)  Disorder operator $\langle e^{i\alpha S^{z}_{A}} \rangle$ as a function of $\alpha$ in the single-replica and two-replica ensembles at $L=768$.
(b) Probability distribution $P(q)$ of the subsystem charge $q$ for $L=64$ and $L=768$.
(c) Scaling of the charge variance $\mathrm{Var}_{q}$ versus $\ln L$.
The red line represents a linear fit, yielding
$\mathrm{Var}_{q}(L)=0.0548(1)\ln L+0.1328(6)$.
(d) Scaling of the number entropy $S^{\mathrm{num}}$ versus $\ln(\ln L)$.
The red line represents a linear fit, yielding
$S^{\mathrm{num}}(L)=0.3483(7)\ln(\ln L)+0.409(1)$.
(e) Finite-size scaling of the subtracted symmetry-resolved R\'enyi entropy, $S_2(q)-S_{A}^{(2)}+\frac{1}{2}\ln(\ln L)$, plotted as a function of $1/\ln L$ for the charge sectors $q=0$ and $q=1$. The lines are linear extrapolations to the thermodynamic limit, $1/\ln L \to 0$.
}
\label{fig:fig4}
\end{figure}

Next, we extend the framework to a system with global $U(1)$ symmetry, using the 1D Heisenberg model as a representative interacting example. We focus on the $U(1)$ subgroup generated by rotations about the $z$ axis, for which the subsystem charge is $Q_A=S_A^z=\sum_{i\in A}S_i^z$, with eigenvalues $q$ ranging from $-L_A/2$ to $L_A/2$. To resolve the individual sectors, we evaluate $\langle e^{i\alpha S_A^z}\rangle$ on a uniform set of counting fields in the single- and two-replica ensembles. All counting fields are accumulated simultaneously from the same configurations, so one simulation in each ensemble provides the complete charged-moment profile; no separate simulation is required for each $\alpha$ or $q$. Spin-inversion symmetry makes the characteristic function real and even, allowing the symmetry-resolved moments to be reconstructed by a discrete cosine transform. The numerical and statistical details are documented in Appendix~\ref{app:U1Reconstruction}.

In $(1+1)$D CFT with a global $U(1)$ symmetry, the SRRE follows a universal scaling behavior. For an equal bipartition of a system with length $L$ and $L_A=L/2$, it is given by~\cite{Estienne_2021}:
\begin{equation}
    S_n(q) = S_{A}^{(n)} - \frac{1}{2} \ln(\ln L) + c_n + \mathcal{O}(1/\ln L),\
    \label{eq:SREE_scaling}
\end{equation}
 which highlights a characteristic subleading double-logarithmic term $-\frac{1}{2} \ln(\ln L)$, alongside strong $q$-dependent finite-size corrections proportional to $\mathcal{O}(1/\ln L)$. 

Equation~\eqref{eq:SREE_scaling} should be understood as equipartition of the asymptotic scaling rather than convergence of the subtracted entropy to a finite common value. Different charge sectors share the sector-independent $-\frac{1}{2}\ln(\ln L)+c_n$ contribution, while their $q$ dependence is relegated to the slowly decaying $\mathcal{O}(1/\ln L)$ finite-size corrections.

Our numerical results are summarized in Fig.~\ref{fig:fig4}. Figure~\ref{fig:fig4}(a) shows the disorder operator $\langle e^{i\alpha S_A^z}\rangle$ measured in the single- and two-replica ensembles at $L=768$. Figures~\ref{fig:fig4}(b) and \ref{fig:fig4}(c) show the subsystem-charge distribution $P(q)$ for $L=64$ and $L=768$, and the charge variance $\mathrm{Var}_q$ as a function of $\ln L$. A linear fit yields
$\mathrm{Var}_{q}(L)=0.0548(1)\ln L+0.1328(6)$,
consistent with the expected logarithmic growth~\cite{PhysRevLett.120.200602}. Figure~\ref{fig:fig4}(d) shows the number entropy,
$S^{\mathrm{num}}=-\sum_q P(q)\ln P(q)$.
A direct fit as a function of $\ln(\ln L)$ gives a slope of $0.3483(7)$, indicating sizable finite-size corrections over the accessible system sizes~\footnote{According to the equipartition prediction, the number entropy in the thermodynamic limit is expected to exhibit the same leading behavior, $S^{\mathrm{num}}\sim \frac{1}{2}\ln(\ln L)$, up to subleading finite-size corrections.}. To reduce these effects, Fig.~\ref{fig:fig4}(e) plots the corrected quantity
$S_n(q)-S_A^{(n)}+\frac{1}{2}\ln(\ln L)$
for the charge sectors $q=0$ and $q=1$ as a function of $1/\ln L$. We focus on these lowest charge sectors because they carry the largest statistical weight and therefore yield the most stable numerical reconstruction. A linear extrapolation to the thermodynamic limit, $1/\ln L\to 0$, gives consistent intercepts for different charge sectors, supporting the scaling form in Eq.~\eqref{eq:SREE_scaling}.


\noindent{\textcolor{blue}{\it Discussion.}---} 

In summary, we have developed a scalable, estimator-based QMC framework for computing symmetry-resolved R\'enyi entropies in sign-problem-free interacting lattice systems. The underlying directed-loop and operator-loop SSE updates are standard; the methodological advance lies in measuring charged moments on glued replica manifolds and reconstructing multiple symmetry sectors from one ordinary-ensemble simulation and one replica-ensemble simulation. This establishes a direct operational connection between field-theoretic charged-moment formulations and large-scale lattice simulations.

The one-dimensional calculations should be viewed as controlled benchmarks of complementary parts of the framework. The 1D TFIM validates the $\mathbb{Z}_2$ replica estimator against an exact CFT prediction, while the interacting Heisenberg chain validates the $U(1)$ symmetry-sector reconstruction and its finite-size scaling. The 2D TFIM illustrates the main intended application beyond one dimension. In the ordinary ensemble, we recover the perimeter-law behavior established in earlier QMC work. In the two-replica ensemble, the accessible data are described by a perimeter law with a subleading term consistent with a logarithmic form, but Eq.~\eqref{eq:2Dfit} remains a phenomenological finite-size parametrization and does not exclude other asymptotic corrections. We have therefore distinguished throughout between directly simulated data and conclusions inferred from extrapolation.

Across the models studied here, the resolved sectors exhibit a common asymptotic entanglement scaling. In the $\mathbb{Z}_2$ TFIM, vanishing nontrivial charged moments imply the common offset $S_2(q)-S_A^{(2)}\to-\ln2$ for the two parity sectors. In the $U(1)$ Heisenberg chain, the charge sectors instead share the same double-logarithmic term and asymptotic intercept, while their differences survive in $\mathcal{O}(1/\ln L)$ corrections. Equipartition thus expresses an emergent sector independence of the leading critical entanglement, while sector probabilities and finite-size corrections retain additional information. Our numerical calculations concern the second R\'enyi entropy and do not by themselves imply identical normalized reduced density matrices or complete entanglement spectra in different sectors, nor do they establish equipartition for all R\'enyi indices.


Access to charged moments on replica manifolds opens several concrete directions. A natural next step is to generalize the method from Abelian symmetries to non-Abelian ones, where symmetry resolution is organized by irreducible representations rather than a single conserved charge. Such an extension would make it possible to study how entanglement is distributed among different representation sectors in interacting lattice models. Another important direction is to apply the method to additional $(2+1)D$ critical systems in order to clarify the structure of subleading corrections to symmetry-resolved entanglement and to test the robustness of entanglement equipartition beyond the Ising universality class. It would also be interesting to use this approach in sign-problem-free models with more exotic ground states, such as interacting topological phases or deconfined quantum critical points, where symmetry-resolved entanglement may provide information that is not accessible from conventional entropy measures alone. More broadly, the method provides a practical numerical tool for studying the interplay between symmetry and entanglement in strongly correlated quantum many-body systems.

\section*{Acknowledgement}
We thank Meng Cheng for valuable discussions on related topics. JRZ and WZJ acknowledge the support from the Research Grants Council (RGC) of Hong Kong Special Administrative Region of China (Project No.~14302725) under the scheme of General Research Fund. JRZ acknowledges the support from  the science panel of Chinese University of Hong Kong (Project No.~4053733) under the Direct Grant scheme.
We thank the Beijing PARATERA Tech CO., Ltd~\cite{paratera} for providing HPC resources that have contributed to the research results reported within this paper.
XPL~acknowledges the Innovation Program for Quantum Science and Technology of China (Grant No.~2024ZD0300100),
the National Basic Research Program of China (Grants No.~2021YFA1400900), 
Shanghai Municipal Science and Technology (Grant No.~25TQ003, 2019SHZDZX01, 24DP2600100). 
RM~acknowledges the support of the German Research Foundation (DFG) through the Collaborative Research Center ToCoTronics, Project-ID 258499086 (SFB 1170), as well as Germany’s Excellence Strategy through the W{\"u}rzburg-Dresden Cluster of Excellence on Complexity and Topology in Quantum Matter - ctd.qmat (EXC 2147, Project-ID 390858490). He furthermore acknowledges hospitality from the Shanghai Institute for Mathematics and Interdisciplinary Sciences (SIMIS), and associated travel support under STCSM Grant 25HB2701900.

\bibliography{references}

@article{Pasquale_Calabrese_2004,
  title = {Entanglement entropy and quantum field theory},
  author = {Calabrese, Pasquale and Cardy, John},
  journal = {J. Stat. Mech.},
  volume = {2004},
  number = {06},
  pages = {P06002},
  year = {2004},
  month = {Jun},
  publisher = {IOP Publishing},
  doi = {10.1088/1742-5468/2004/06/P06002},
  url = {http://dx.doi.org/10.1088/1742-5468/2004/06/P06002}
}

@article{zhao2022charged,
  title={Charged moments in W3 higher spin holography},
  author={Zhao, Suting and Northe, Christian and Weisenberger, Konstantin and Meyer, Ren{\'e}},
  journal={Journal of High Energy Physics},
  volume={2022},
  number={5},
  pages={166},
  year={2022},
  doi={10.1007/JHEP05(2022)166},
  url={https://doi.org/10.1007/JHEP05(2022)166},
  publisher={Springer}
}

@article{RevModPhys.80.517,
  title = {Entanglement in many-body systems},
  author = {Amico, Luigi and Fazio, Rosario and Osterloh, Andreas and Vedral, Vlatko},
  journal = {Rev. Mod. Phys.},
  volume = {80},
  issue = {2},
  pages = {517--576},
  numpages = {0},
  year = {2008},
  month = {May},
  publisher = {American Physical Society},
  doi = {10.1103/RevModPhys.80.517},
  url = {https://link.aps.org/doi/10.1103/RevModPhys.80.517}
}

@article{Laflorencie_2016,
  title = {Quantum entanglement in condensed matter systems},
  author = {Laflorencie, Nicolas},
  journal = {Phys. Rep.},
  volume = {646},
  pages = {1--59},
  year = {2016},
  month = {Aug},
  publisher = {Elsevier},
  doi = {10.1016/j.physrep.2016.06.008},
  url = {http://dx.doi.org/10.1016/j.physrep.2016.06.008}
}

@article{Parez2021Exact,
    author = {Parez, Gilles and Bonsignori, Riccarda and Calabrese, Pasquale},
    title = {Exact quench dynamics of symmetry resolved entanglement in a free fermion chain},
    primaryClass = {cond-mat.stat-mech},
    doi = {10.1088/1742-5468/ac21d7},
    journal = {J. Stat. Mech.},
    volume = {2021},
    number = {9},
    pages = {093102},
    year = {2021},
}

@Article{10.21468/SciPostPhys.17.5.127,
	title={{Non-Abelian symmetry-resolved entanglement entropy}},
	author={Eugenio Bianchi and Pietro Dona and Rishabh Kumar},
	journal={SciPost Phys.},
	volume={17},
	pages={127},
	year={2024},
	publisher={SciPost},
	doi={10.21468/SciPostPhys.17.5.127},
	url={https://scipost.org/10.21468/SciPostPhys.17.5.127},
}

@article{belin2013holographic,
  title={Holographic charged R{\'e}nyi entropies},
  author={Belin, Alexandre and Hung, Ling-Yan and Maloney, Alexander and Matsuura, Shunji and Myers, Robert C and Sierens, Todd},
  journal={Journal of High Energy Physics},
  volume={2013},
  number={12},
  pages={1--50},
  year={2013},
  publisher={Springer},
  doi={10.1007/JHEP12(2013)059},
  url={https://doi.org/10.1007/JHEP12(2013)059}
}

@article{PhysRevLett.120.200602,
  title = {Symmetry-Resolved Entanglement in Many-Body Systems},
  author = {Goldstein, Moshe and Sela, Eran},
  journal = {Phys. Rev. Lett.},
  volume = {120},
  issue = {20},
  pages = {200602},
  numpages = {6},
  year = {2018},
  month = {May},
  publisher = {American Physical Society},
  doi = {10.1103/PhysRevLett.120.200602},
  url = {https://link.aps.org/doi/10.1103/PhysRevLett.120.200602}
}

@article{Laflorencie_2014,
  title = {Spin-resolved entanglement spectroscopy of critical spin chains and Luttinger liquids},
  author = {Laflorencie, Nicolas and Rachel, Stephan},
  journal = {J. Stat. Mech.},
  volume = {2014},
  number = {11},
  pages = {P11013},
  year = {2014},
  month = {Nov},
  publisher = {IOP Publishing},
  doi = {10.1088/1742-5468/2014/11/P11013},
  url = {http://dx.doi.org/10.1088/1742-5468/2014/11/P11013}
}

@article{RevModPhys.82.277,
  title = {Colloquium: Area laws for the entanglement entropy},
  author = {Eisert, J. and Cramer, M. and Plenio, M. B.},
  journal = {Rev. Mod. Phys.},
  volume = {82},
  issue = {1},
  pages = {277--306},
  numpages = {0},
  year = {2010},
  month = {Feb},
  publisher = {American Physical Society},
  doi = {10.1103/RevModPhys.82.277},
  url = {https://link.aps.org/doi/10.1103/RevModPhys.82.277}
}

@article{PhysRevB.98.041106,
  title = {Equipartition of the entanglement entropy},
  author = {Xavier, J. C. and Alcaraz, F. C. and Sierra, G.},
  journal = {Phys. Rev. B},
  volume = {98},
  issue = {4},
  pages = {041106},
  numpages = {6},
  year = {2018},
  month = {Jul},
  publisher = {American Physical Society},
  doi = {10.1103/PhysRevB.98.041106},
  url = {https://link.aps.org/doi/10.1103/PhysRevB.98.041106}
}

@article{Estienne_2021,
  title = {Finite-size corrections in critical symmetry-resolved entanglement},
  author = {Estienne, Benoit and Ikhlef, Yacine and Morin-Duchesne, Alexi},
  journal = {SciPost Phys.},
  volume = {10},
  number = {3},
  pages = {054},
  year = {2021},
  month = {Mar},
  doi = {10.21468/SciPostPhys.10.3.054},
  url = {http://dx.doi.org/10.21468/SciPostPhys.10.3.054}
}

@article{Bonsignori_2019,
  title = {Symmetry resolved entanglement in free fermionic systems},
  author = {Bonsignori, Riccarda and Ruggiero, Paola and Calabrese, Pasquale},
  journal = {J. Phys. A: Math. Theor.},
  volume = {52},
  number = {47},
  pages = {475302},
  year = {2019},
  month = {Oct},
  publisher = {IOP Publishing},
  doi = {10.1088/1751-8121/ab4b77},
  url = {http://dx.doi.org/10.1088/1751-8121/ab4b77}
}

@article{calabrese2021symmetry,
  title={Symmetry-resolved entanglement entropy in Wess-Zumino-Witten models},
  author={Calabrese, Pasquale and Dubail, J{\'e}r{\^o}me and Murciano, Sara},
  journal={Journal of High Energy Physics},
  volume={2021},
  number={10},
  pages={67},
  year={2021},
  publisher={Springer},
  doi = {10.1007/JHEP10(2021)067},
  url={https://doi.org/10.1007/JHEP10(2021)067}
}

@article{Capizzi_2020,
  title = {Symmetry resolved entanglement entropy of excited states in a CFT},
  author = {Capizzi, Luca and Ruggiero, Paola and Calabrese, Pasquale},
  journal = {J. Stat. Mech.},
  volume = {2020},
  number = {07},
  pages = {073101},
  year = {2020},
  month = {Jul},
  publisher = {IOP Publishing},
  doi = {10.1088/1742-5468/ab96b6},
  url = {http://dx.doi.org/10.1088/1742-5468/ab96b6}
}

@article{zhao2021symmetry,
  title={Symmetry-resolved entanglement in AdS3/CFT2 coupled to U (1) Chern-Simons theory},
  author={Zhao, Suting and Northe, Christian and Meyer, Ren{\'e}},
  journal={Journal of High Energy Physics},
  volume={2021},
  number={7},
  pages={30},
  year={2021},
  doi={10.1007/JHEP07(2021)030},
  url={https://doi.org/10.1007/JHEP07(2021)030},
  publisher={Springer}
}

@article{weisenberger2021symmetry,
  title={Symmetry-resolved entanglement for excited states and two entangling intervals in AdS3/CFT2},
  author={Weisenberger, Konstantin and Zhao, Suting and Northe, Christian and Meyer, Ren{\'e}},
  journal={Journal of High Energy Physics},
  volume={2021},
  number={12},
  pages={104},
  year={2021},
  doi={10.1007/JHEP12(2021)104},
  url={https://doi.org/10.1007/JHEP12(2021)104},
  publisher={Springer}
}

@article{digiulio2023boundaryconformalfieldtheory,
  title={On the boundary conformal field theory approach to symmetry-resolved entanglement},
  author={Di Giulio, Giuseppe and Meyer, Ren{\'e} and Northe, Christian and Scheppach, Henri and Zhao, Suting},
  journal={SciPost Physics Core},
  volume={6},
  number={3},
  pages={049},
  year={2023},
  doi = {10.21468/SciPostPhysCore.6.3.049},
  url = {https://scipost.org/10.21468/SciPostPhysCore.6.3.049}
}

@article{huang2025symmetryresolvedentanglemententropyhigher,
  title={Symmetry-resolved entanglement entropy in higher dimensions},
  author={Huang, Yuanzhu and Zhou, Yang},
  journal={Journal of High Energy Physics},
  volume={2025},
  number={10},
  pages={1--32},
  year={2025},
  publisher={Springer},
  doi={10.1007/JHEP10(2025)045},
  url={https://doi.org/10.1007/JHEP10(2025)045}
}

@article{Murciano_2020,
  title = {Symmetry resolved entanglement in two-dimensional systems via dimensional reduction},
  author = {Murciano, Sara and Ruggiero, Paola and Calabrese, Pasquale},
  journal = {J. Stat. Mech.},
  volume = {2020},
  number = {08},
  pages = {083102},
  year = {2020},
  month = {Aug},
  publisher = {IOP Publishing},
  doi = {10.1088/1742-5468/aba1e5},
  url = {http://dx.doi.org/10.1088/1742-5468/aba1e5}
}

@article{Lukin_2019,
  title = {Probing entanglement in a many-body--localized system},
  author = {Lukin, Alexander and Rispoli, Matthew and Schittko, Robert and Tai, M. Eric and Kaufman, Adam M. and Choi, Soonwon and Khemani, Vedika and L{\'e}onard, Julian and Greiner, Markus},
  journal = {Science},
  volume = {364},
  number = {6437},
  pages = {256--260},
  year = {2019},
  month = {Apr},
  publisher = {American Association for the Advancement of Science},
  doi = {10.1126/science.aau0818},
  url = {http://dx.doi.org/10.1126/science.aau0818}
}

@article{bueno2022universal,
    author = "Bueno, Pablo and Cano, Pablo A. and Murcia, {\'A}ngel and Rivadulla S{\'a}nchez, Alberto",
    title = "Universal Feature of Charged Entanglement Entropy",
    reportNumber = "IFT-UAM/CSIC-22-18, CERN-TH-2022-033",
    doi = "10.1103/PhysRevLett.129.021601",
    journal = "Phys. Rev. Lett.",
    volume = "129",
    number = "2",
    pages = "021601",
    year = "2022"
}

@article{Bloete2002,
	title = {Cluster Monte Carlo simulation of the transverse Ising model},
	author = {Bl\"ote, Henk W. J. and Deng, Youjin},
	journal = {Phys. Rev. E},
	volume = {66},
	issue = {6},
	pages = {066110},
	numpages = {8},
	year = {2002},
	month = {Dec},
	publisher = {American Physical Society},
	doi = {10.1103/PhysRevE.66.066110},
	url = {https://link.aps.org/doi/10.1103/PhysRevE.66.066110}
}

@article{PanGaopei2023,
  title = {Stable computation of entanglement entropy for two-dimensional interacting fermion systems},
  author = {Pan, Gaopei and Da Liao, Yuan and Jiang, Weilun and D'Emidio, Jonathan and Qi, Yang and Meng, Zi Yang},
  journal = {Phys. Rev. B},
  volume = {108},
  issue = {8},
  pages = {L081123},
  numpages = {6},
  year = {2023},
  month = {Aug},
  publisher = {American Physical Society},
  doi = {10.1103/PhysRevB.108.L081123},
  url = {https://link.aps.org/doi/10.1103/PhysRevB.108.L081123}
}

@article{EmidioUniversal2024,
  title = {Universal Features of Entanglement Entropy in the Honeycomb Hubbard Model},
  author = {D'Emidio, Jonathan and Or\'us, Rom\'an and Laflorencie, Nicolas and de Juan, Fernando},
  journal = {Phys. Rev. Lett.},
  volume = {132},
  issue = {7},
  pages = {076502},
  numpages = {6},
  year = {2024},
  month = {Feb},
  publisher = {American Physical Society},
  doi = {10.1103/PhysRevLett.132.076502},
  url = {https://link.aps.org/doi/10.1103/PhysRevLett.132.076502}
}

@article{ZihongItinerant2019,
	author = {Liu, Zi Hong and Pan, Gaopei and Xu, Xiao Yan and Sun, Kai and Meng, Zi Yang},
	title = {Itinerant quantum critical point with fermion pockets and hotspots},
	volume = {116},
	number = {34},
	pages = {16760--16767},
	year = {2019},
	doi = {10.1073/pnas.1901751116},
	publisher = {National Academy of Sciences},
	issn = {0027-8424},
	URL = {https://www.pnas.org/content/116/34/16760},
	journal = {Proc. Natl. Acad. Sci. U.S.A.}
}

@ARTICLE{zhaoHigher2021,
  title = {Higher-form symmetry breaking at Ising transitions},
  author = {Zhao, Jiarui and Yan, Zheng and Cheng, Meng and Meng, Zi Yang},
  journal = {Phys. Rev. Res.},
  volume = {3},
  issue = {3},
  pages = {033024},
  numpages = {12},
  year = {2021},
  month = {Jul},
  publisher = {American Physical Society},
  doi = {10.1103/PhysRevResearch.3.033024},
  url = {https://link.aps.org/doi/10.1103/PhysRevResearch.3.033024}
}

@article{wangScaling2021,
  title = {Scaling of the disorder operator at $(2+1)d$ U(1) quantum criticality},
  author = {Wang, Yan-Cheng and Cheng, Meng and Meng, Zi Yang},
  journal = {Phys. Rev. B},
  volume = {104},
  issue = {8},
  pages = {L081109},
  numpages = {5},
  year = {2021},
  month = {Aug},
  publisher = {American Physical Society},
  doi = {10.1103/PhysRevB.104.L081109},
  url = {https://link.aps.org/doi/10.1103/PhysRevB.104.L081109}
}

@article{zhaoScaling2022,
  title = {Scaling of Entanglement Entropy at Deconfined Quantum Criticality},
  author = {Zhao, Jiarui and Wang, Yan-Cheng and Yan, Zheng and Cheng, Meng and Meng, Zi Yang},
  journal = {Phys. Rev. Lett.},
  volume = {128},
  issue = {1},
  pages = {010601},
  numpages = {6},
  year = {2022},
  month = {Jan},
  publisher = {American Physical Society},
  doi = {10.1103/PhysRevLett.128.010601},
  url = {https://link.aps.org/doi/10.1103/PhysRevLett.128.010601}
}

@article{zhaoMeasuring2022,
    author = "Zhao, Jiarui and Chen, Bin-Bin and Wang, Yan-Cheng and Yan, Zheng and Cheng, Meng and Meng, Zi Yang",
    title = "{Measuring R{\'e}nyi entanglement entropy with high efficiency and precision in quantum Monte Carlo simulations}",
    primaryClass = "cond-mat.str-el",
    doi = "10.1038/s41535-022-00476-0",
    journal = "npj Quant. Mater.",
    volume = "7",
    number = "1",
    pages = "69",
    year = "2022"
}

@article{song2023deconfined,
    author = "Song, Menghan and Zhao, Jiarui and Cheng, Meng and Xu, Cenke and Scherer, Michael M. and Janssen, Lukas and Yang Meng, Zi",
    title = "{Evolution of entanglement entropy at SU(N) deconfined quantum critical points}",
    doi = "10.1126/sciadv.adr0634",
    journal = "Sci. Adv.",
    volume = "11",
    number = "6",
    pages = "adr0634",
    year = "2025"
}

@article{Sandvik_2002,
  title = {Quantum Monte Carlo with directed loops},
  author = {Sylju\aa{}sen, Olav F. and Sandvik, Anders W.},
  journal = {Phys. Rev. E},
  volume = {66},
  issue = {4},
  pages = {046701},
  numpages = {28},
  year = {2002},
  month = {Oct},
  publisher = {American Physical Society},
  doi = {10.1103/PhysRevE.66.046701},
  url = {https://link.aps.org/doi/10.1103/PhysRevE.66.046701}
}

@Article{wangScaling2022,
	title={{Scaling of the disorder operator at deconfined quantum criticality}},
	author={Yan-Cheng Wang and Nvsen Ma and Meng Cheng and Zi Yang Meng},
	journal={SciPost Phys.},
	volume={13},
	pages={123},
	year={2022},
	publisher={SciPost},
	doi={10.21468/SciPostPhys.13.6.123},
	url={https://scipost.org/10.21468/SciPostPhys.13.6.123},
}

@article{kulchytskyyDetecting2015,
  title = {Detecting Goldstone modes with entanglement entropy},
  author = {Kulchytskyy, Bohdan and Herdman, C. M. and Inglis, Stephen and Melko, Roger G.},
  journal = {Phys. Rev. B},
  volume = {92},
  issue = {11},
  pages = {115146},
  numpages = {11},
  year = {2015},
  month = {Sep},
  publisher = {American Physical Society},
  doi = {10.1103/PhysRevB.92.115146},
  url = {https://link.aps.org/doi/10.1103/PhysRevB.92.115146}
}

@article{Sandvik1999,
	title = {Stochastic series expansion method with operator-loop update},
	author = {Sandvik, Anders W.},
	journal = {Phys. Rev. B},
	volume = {59},
	issue = {22},
	pages = {R14157--R14160},
	numpages = {0},
	year = {1999},
	month = {Jun},
	publisher = {American Physical Society},
	doi = {10.1103/PhysRevB.59.R14157},
	url = {https://link.aps.org/doi/10.1103/PhysRevB.59.R14157}
}

@article{IsakovTopological2011,
  author = {Isakov, Sergei V. and Hastings, Matthew B. and Melko, Roger G.},
  title = {Topological entanglement entropy of a Bose–Hubbard spin liquid},
  journal = {Nature Phys.},
  volume = {7},
  number = {10},
  pages = {772--775},
  year = {2011},
  doi = {10.1038/nphys2036},
  url = {https://doi.org/10.1038/nphys2036},
  issn = {1745-2481},
}

@article{demidioEntanglement2020,
  title = {Entanglement Entropy from Nonequilibrium Work},
  author = {D'Emidio, Jonathan},
  journal = {Phys. Rev. Lett.},
  volume = {124},
  issue = {11},
  pages = {110602},
  numpages = {5},
  year = {2020},
  month = {Mar},
  publisher = {American Physical Society},
  doi = {10.1103/PhysRevLett.124.110602},
  url = {https://link.aps.org/doi/10.1103/PhysRevLett.124.110602}
}

@article{Jarzynski1997,
	title = {Nonequilibrium Equality for Free Energy Differences},
	author = {Jarzynski, C.},
	journal = {Phys. Rev. Lett.},
	volume = {78},
	issue = {14},
	pages = {2690--2693},
	numpages = {0},
	year = {1997},
	month = {Apr},
	publisher = {American Physical Society},
	doi = {10.1103/PhysRevLett.78.2690},
	url = {https://link.aps.org/doi/10.1103/PhysRevLett.78.2690}
}

@article{kitaevTopological2006,
	title = {Topological Entanglement Entropy},
	author = {Kitaev, Alexei and Preskill, John},
	journal = {Phys. Rev. Lett.},
	volume = {96},
	issue = {11},
	pages = {110404},
	numpages = {4},
	year = {2006},
	month = {Mar},
	publisher = {American Physical Society},
	doi = {10.1103/PhysRevLett.96.110404},
	url = {https://link.aps.org/doi/10.1103/PhysRevLett.96.110404}
}

@misc{paratera,
  author       = {{Beijing PARATERA Tech Co., Ltd.}},
  title        = {PARATERA Cloud Platform},
  url          = {https://cloud.paratera.com},
}

@article{SkinnerPRX2019,
  title = {Measurement-Induced Phase Transitions in the Dynamics of Entanglement},
  author = {Skinner, Brian and Ruhman, Jonathan and Nahum, Adam},
  journal = {Phys. Rev. X},
  volume = {9},
  issue = {3},
  pages = {031009},
  numpages = {21},
  year = {2019},
  month = {Jul},
  publisher = {American Physical Society},
  doi = {10.1103/PhysRevX.9.031009},
  url = {https://link.aps.org/doi/10.1103/PhysRevX.9.031009}
}

@Article{ROeland2023,
	title={{Measurement-induced entanglement phase transitions in variational quantum circuits}},
	author={Roeland Wiersema and Cunlu Zhou and Juan Felipe Carrasquilla and Yong Baek Kim},
	journal={SciPost Phys.},
	volume={14},
	pages={147},
	year={2023},
	publisher={SciPost},
	doi={10.21468/SciPostPhys.14.6.147},
	url={https://scipost.org/10.21468/SciPostPhys.14.6.147},
}

@article{Paviglianiti2023ee,
  title = {Multipartite entanglement in the measurement-induced phase transition of the quantum Ising chain},
  author = {Paviglianiti, Alessio and Silva, Alessandro},
  journal = {Phys. Rev. B},
  volume = {108},
  issue = {18},
  pages = {184302},
  numpages = {7},
  year = {2023},
  month = {Nov},
  publisher = {American Physical Society},
  doi = {10.1103/PhysRevB.108.184302},
  url = {https://link.aps.org/doi/10.1103/PhysRevB.108.184302}
}

@misc{jain2025,
  author        = {Jain, Ashwat},
  title         = {Symmetry Resolved Multipartite Entanglement Entropy},
  year          = {2025},
  month         = sep,
  eprint        = {2509.13410},
  archivePrefix = {arXiv},
  primaryClass  = {quant-ph}
}

@article{SCHOLLWOCK201196,
    author = "Schollwoeck, Ulrich",
    title = "{The density-matrix renormalization group in the age of matrix product states}",
    doi = "10.1016/j.aop.2010.09.012",
    journal = "Annals Phys.",
    volume = "326",
    pages = "96--192",
    year = "2011"
}

@misc{Chen2026Renyi,
  title  = {Scaling of {R{\'e}nyi} Disorder Operators at {(2+1)D} Conformal Quantum Criticality},
  author = {Kuangjie Chen and Weizhen Jia and Xiaopeng Li and Ren\'e Meyer and Meng Cheng and Jiarui Zhao},
  note   = {to appear},
}

@article{FraenkelGoldstein2020,
  title = {Symmetry resolved entanglement: Exact results in 1D and beyond},
  author = {Fraenkel, Shachar and Goldstein, Moshe},
  journal = {J. Stat. Mech.},
  volume = {2020},
  number = {03},
  pages = {033106},
  year = {2020},
  month = {Mar},
  doi = {10.1088/1742-5468/ab7753},
  url = {https://doi.org/10.1088/1742-5468/ab7753}
}

@article{TanRyu2020,
  title = {Particle number fluctuations, {R{\'e}nyi} entropy, and symmetry-resolved entanglement entropy in a two-dimensional Fermi gas from multidimensional bosonization},
  author = {Tan, Mao Tian and Ryu, Shinsei},
  journal = {Phys. Rev. B},
  volume = {101},
  issue = {23},
  pages = {235169},
  year = {2020},
  month = {Jun},
  publisher = {American Physical Society},
  doi = {10.1103/PhysRevB.101.235169},
  url = {https://doi.org/10.1103/PhysRevB.101.235169}
}

@article{XuHuang2025,
  title = {Finite Correlation Length Scaling of Disorder Parameter at Quantum Criticality},
  author = {Xu, Wen-Tao and Huang, Rui-Zhen},
  journal = {Phys. Rev. Lett.},
  volume = {134},
  issue = {14},
  pages = {146503},
  year = {2025},
  month = {Apr},
  publisher = {American Physical Society},
  doi = {10.1103/PhysRevLett.134.146503},
  url = {https://doi.org/10.1103/PhysRevLett.134.146503}
}

@article{MonkmanSirker2023General,
  title = {Symmetry-resolved entanglement: general considerations, calculation from correlation functions, and bounds for symmetry-protected topological phases},
  author = {Monkman, Kyle and Sirker, Jesko},
  journal = {J. Phys. A: Math. Theor.},
  volume = {56},
  number = {49},
  pages = {495001},
  numpages = {20},
  year = {2023},
  month = {Nov},
  publisher = {IOP Publishing},
  doi = {10.1088/1751-8121/ad086d},
  url = {https://doi.org/10.1088/1751-8121/ad086d}
}

@article{PhysRevB.107.125108,
  title = {Symmetry-resolved entanglement of {$C_2$}-symmetric topological insulators},
  author = {Monkman, Kyle and Sirker, Jesko},
  journal = {Phys. Rev. B},
  volume = {107},
  issue = {12},
  pages = {125108},
  year = {2023},
  month = {Mar},
  publisher = {American Physical Society},
  doi = {10.1103/PhysRevB.107.125108},
  url = {https://link.aps.org/doi/10.1103/PhysRevB.107.125108}
}

@article{PhysRevResearch.2.043191,
  title = {Operational entanglement of symmetry-protected topological edge states},
  author = {Monkman, Kyle and Sirker, Jesko},
  journal = {Phys. Rev. Research},
  volume = {2},
  issue = {4},
  pages = {043191},
  numpages = {9},
  year = {2020},
  month = {Nov},
  publisher = {American Physical Society},
  doi = {10.1103/PhysRevResearch.2.043191},
  url = {https://link.aps.org/doi/10.1103/PhysRevResearch.2.043191}
}

\clearpage
\onecolumngrid
\appendix

\section{Benchmark of QMC Results}
\label{app:Benchmark}
To verify the validity and accuracy of our stochastic series expansion QMC implementations, we first compare the results of $\langle X \rangle_Z$ and $\langle X \rangle_{Z_A^{(2)}}$ obtained from QMC simulations at extremely low temperatures with those derived from exact diagonalization (ED) at zero temperature. The comparison is performed on both the 1D and 2D Ising models at their respective critical points. 

For the 1D Ising model ($h_c = 1$), we compare the ED results with our QMC results at $\beta = 256$ and $L = 16$, where subsystem $A$ is defined as half of the periodic chain. For the 2D Ising model ($h_c \approx 3.044$), we compare the ED results with the QMC results at $L_x = L_y = 4$ and $\beta = 128$. Here, subsystem $A$ is chosen as an $(L_x/2)\times L_y$ cylinder on an $L_x \times L_y$ torus. As shown in Table~\ref{benchmark}, our QMC results agree well with the ED results within statistical errors for both the 1D and 2D Ising critical points.

\begin{table}[htbp]
  \centering
  \caption{Comparison of ED and QMC results at the 1D and 2D Ising critical points.}
  \begin{tabular}{llcc}
    \toprule
    & & $\langle X\rangle_{Z}$ & $\langle X\rangle_{Z_A^{(2)}}$ \\
    \midrule
    \multirow{2}{*}{1D $\quad$}
      & ED $\quad$  & 0.429185 $\quad$  & 0.761487 \\
      & QMC $\quad$ & 0.4293(2) $\quad$ & 0.7616(1)  \\
    \midrule
    \multirow{2}{*}{2D $\quad$}
      & ED $\quad$ & 0.7007641 $\quad$ & 0.9556409\\
      & QMC $\quad$ & 0.70072(5) $\quad$ & 0.95566(2) \\
    \bottomrule
  \end{tabular}
  \label{benchmark}
\end{table}

\begin{figure}[htp]
  \centering
  \includegraphics[width=\columnwidth]{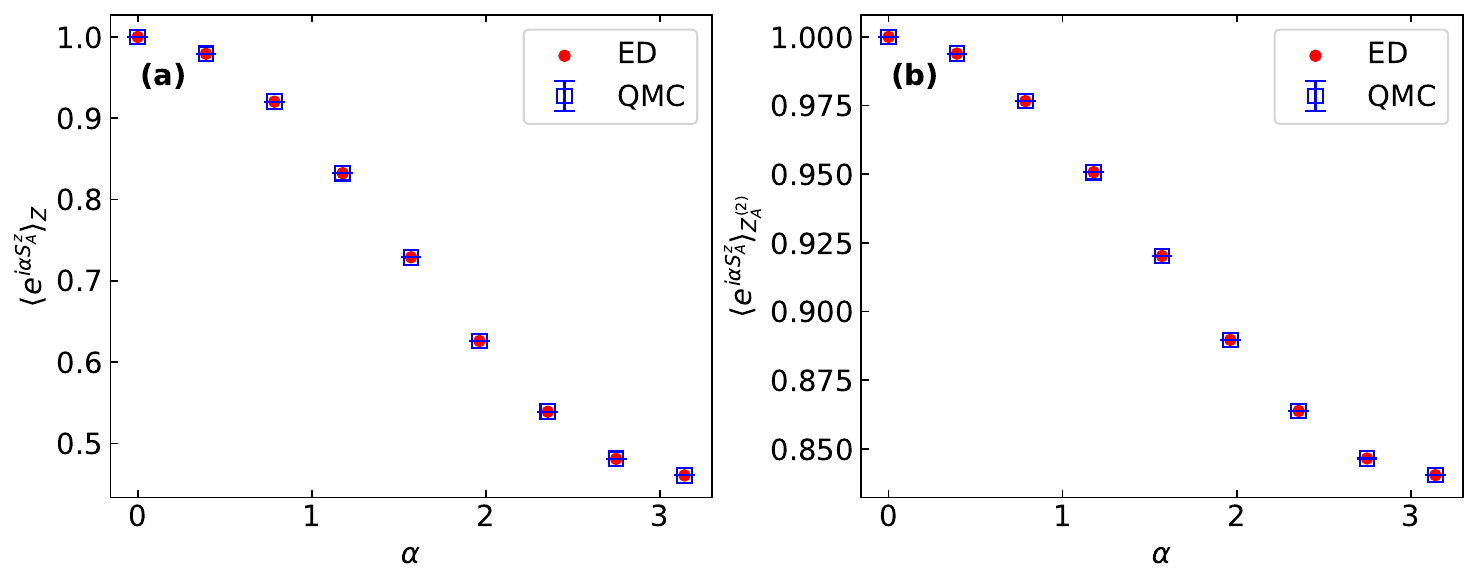}
  \caption{Comparison of ED and QMC results for the disorder operator in the 1D Heisenberg model.}
  \label{fig:figS2}
\end{figure}

We further analyze the disorder operator $\langle e^{i \alpha S^{z}_{A}} \rangle_{Z}$ and $\langle e^{i \alpha S^{z}_{A}} \rangle_{Z_A^{(2)}}$ for the 1D Heisenberg model. Comparing the ED results with our QMC data at $\beta = 256$ and $L = 16$, we observe excellent agreement between the two methods, as illustrated in Fig.~\ref{fig:figS2}.

\section{Numerical Implementation, Symmetry-Sector Reconstruction, and Error Analysis}
\label{app:U1Reconstruction}

\subsection{SSE sampling and replica geometry}

This appendix summarizes the numerical implementation in the ordinary and replica ensembles and then details the reconstruction of the $U(1)$-resolved second R\'enyi entropy for the spin-$1/2$ Heisenberg chain. For the TFIM, we work in the $\sigma^x$ basis and use directed-loop updates, so that the $\mathbb{Z}_2$ twist $X=\prod_{i\in A}\sigma_i^x$ is diagonal in the simulation basis. For the Heisenberg chain, we work in the $S^z$ basis and use operator-loop updates, with the subsystem charge $Q_A=S_A^z$ measured directly. In both models, the ordinary and two-replica ensembles use the same model-specific update machinery. The distinction lies in the imaginary-time boundary connectivity: degrees of freedom in region $A$ connect between the two replicas, whereas those in $\bar{A}$ close within each replica. No independent Monte Carlo update or simulation is required for each counting field or symmetry sector.


During our QMC simulations, the inverse temperature is set to be $\beta = 2L $ for the Heisenberg chain.  For each system size, the single-replica and two-replica ensembles were simulated separately. Each ensemble comprised 256 independent rank-level Monte Carlo streams, and each stream was equilibrated for $10^5$ sweeps before measurements. The production data for each ensemble were accumulated into a total of 1,280 statistical bins, corresponding to five bins per stream, with each bin containing $10^5$ measurement sweeps. Thus, for each system size, $1.28\times10^8$ production sweeps were performed in each ensemble. Observables were accumulated at every measurement sweeps.

\subsection{$U(1)$ charged moments and discrete reconstruction}

We consider a periodic Heisenberg chain of length $L$ and choose the entangling region $A$ to be a half-chain of length $L_A=L/2$. The local convention is $S_i^z=\pm1/2$, and the physical subsystem charge is the total magnetization, $Q_A=S_A^z$. Since all system sizes considered in our simulations are multiples of four, $L_A$ is even and the allowed subsystem charges are integers,

\begin{equation}
    q=-q_{\max},\ldots,q_{\max},
    \qquad
    q_{\max}=\frac{L_A}{2}=\frac{L}{4}.
    \label{eq:app_charge_range}
\end{equation}

We perform independent SSE simulations in the ordinary ensemble $Z$ and the two-replica ensemble $Z_A^{(2)}$. The corresponding charged moments are
\begin{align}
    \chi_1(\alpha)
    &=
    \left\langle e^{i\alpha Q_A}\right\rangle_Z
    =
    \frac{
    \operatorname{Tr}\!\left(\rho_Ae^{i\alpha Q_A}\right)
    }{
    \operatorname{Tr}\rho_A
    },
    \label{eq:app_chi1}\\
    \chi_2(\alpha)
    &=
    \left\langle e^{i\alpha Q_A}\right\rangle_{Z_A^{(2)}}
    =
    \frac{
    \operatorname{Tr}\!\left(\rho_A^2e^{i\alpha Q_A}\right)
    }{
    \operatorname{Tr}\rho_A^2
    }.
    \label{eq:app_chi2}
\end{align}
Spin-inversion symmetry implies that the charge distributions are symmetric under $q\rightarrow-q$. The charged moments are therefore real and even functions of $\alpha$, and can be evaluated through their real parts,
\begin{equation}
    \chi_n(\alpha)
    =
    \left\langle\cos(\alpha Q_A)\right\rangle.
\end{equation}

Since $Q_A$ is diagonal in the $S^z$ basis used in the SSE simulation, its measurement requires neither an additional Monte Carlo update nor an independent simulation for each value of $\alpha$. For each measured configuration, $Q_A$ is evaluated once and all quantities $\cos(\alpha_jQ_A)$ are calculated simultaneously. One simulation in the ordinary ensemble and one in the two-replica ensemble therefore provide the complete charged-moment profiles for all counting fields. No separate simulation is required for each value of $\alpha$ or for each charge sector.

Defining $M=L_A=L/2$, we evaluate the charged moments on the uniform grid
\begin{equation}
    \alpha_j=\frac{j\pi}{M},
    \qquad
    j=0,\ldots,M.
    \label{eq:app_alpha_grid}
\end{equation}
The values at $j=1,\ldots,M$ are measured directly in the Monte Carlo simulation, while the endpoint at $j=0$ is fixed exactly as $\chi_1(0)=\chi_2(0)=1$. Since the spectrum of $Q_A$ has finite integer support, $|q|\leq M/2$, and the charge distributions are symmetric under $q\rightarrow-q$, the charged moments are finite cosine polynomials of degree at most $M/2$. The weighted DCT-I formula on the grid $\alpha_j=j\pi/M$, $j=0,\ldots,M$, therefore exactly reconstructs every physical charge sector $0\leq q\leq M/2$. Consequently, for noise-free charged moments, the discrete reconstruction is identical to the continuous Fourier transform up to floating-point roundoff and introduces neither aliasing nor Fourier-truncation errors. We have also explicitly verified that the reconstructed results remain unchanged within statistical uncertainties when denser $\alpha$ grids are used.

Using the symmetry $P(q)=P(-q)$, the ordinary charge distribution is reconstructed through the discrete cosine sum
\begin{equation}
    P(q)
    =
    \frac{1}{M}
    \bigg[
       \frac{1}{2}\chi_1(0)
       +\sum_{j=1}^{M-1}\chi_1(\alpha_j)\cos(q\alpha_j)
       +\frac{1}{2}\chi_1(\pi)\cos(q\pi)
    \bigg],
    \qquad
    0\leq q\leq\frac{M}{2}.
    \label{eq:app_reconstruct_p}
\end{equation}
The factors of $1/2$ at the two endpoints are the standard trapezoidal, or equivalently discrete-cosine-transform type-I, weights. The negative-charge sectors are obtained from $P(-q)=P(q)$.

The same transformation is applied to the normalized two-replica charged moment:
\begin{equation}
    R_2(q)
    =
    \frac{1}{M}
    \bigg[
       \frac{1}{2}\chi_2(0)
       +\sum_{j=1}^{M-1}\chi_2(\alpha_j)\cos(q\alpha_j)
       +\frac{1}{2}\chi_2(\pi)\cos(q\pi)
    \bigg].
    \label{eq:app_reconstruct_r2}
\end{equation}
Here,
\begin{equation}
    R_2(q)
    =
    \frac{
    \operatorname{Tr}(\Pi_q\rho_A^2)
    }{
    \operatorname{Tr}\rho_A^2
    }.
    \label{eq:app_r2_definition}
\end{equation}
The subtracted symmetry-resolved second R\'enyi entropy is then obtained from
\begin{equation}
    S_2(q)-S_A^{(2)}
    =
    -\ln
    \left[
    \frac{R_2(q)}{P(q)^2}
    \right].
    \label{eq:app_sree_reconstruction}
\end{equation}

\subsection{Statistical analysis, stability, and computational cost}

The reconstruction is a direct linear cosine transform rather than an ill-conditioned inversion or analytic continuation. As internal stability checks, we monitor the normalization $\sum_{q=-q_{\max}}^{q_{\max}}P(q)=1$ and verify that a denser counting-field grid leaves the reconstructed results unchanged within statistical uncertainties. Statistical noise becomes increasingly important in the tails of the charge distribution, where the sector weights are very small. We therefore restrict the SRRE analysis and finite-size extrapolations to the dominant $q=0$ and $q=1$ sectors, for which the reconstructed $P(q)$ and $R_2(q)$ are positive and statistically well resolved.


The Monte Carlo measurements were accumulated into predefined bins, with
each bin containing the complete charged-moment vector at the
\(M=L/2\) measured counting fields. No additional rebinning was performed
during post-processing. We applied the DCT-I reconstruction separately to
each bin, thereby obtaining bin-level estimates of \(P(q)\) and
\(R_2(q)\). The marginal uncertainties in the reconstructed \(P(q)\) and
\(R_2(q)\) distributions were evaluated as standard errors across the
corresponding bin-level DCT-I estimates. Since the complete counting-field
vector was transformed jointly within each bin, this procedure retains the
covariance among the different \(\alpha_j\) values measured from the same
Monte Carlo configurations.

For the nonlinear quantity \(S_2(q)-S_A^{(2)}\), we employed an independent
two-sample nonparametric bootstrap with \(10^{4}\) resamples. In each
bootstrap realization, the bins from the ordinary and two-replica ensembles
were resampled independently with replacement, while retaining the original
number of bins in each ensemble. All components associated with a selected
bin were kept together during the resampling. The resulting bootstrap means,
\(\overline{P}^{\,*}(q)\) and \(\overline{R}_2^{\,*}(q)\), were then used
to evaluate
\begin{equation}
    \Delta S_2^{*}(q)
    =
    -\ln\!\left[
        \frac{\overline{R}_2^{\,*}(q)}
        {\left[\overline{P}^{\,*}(q)\right]^2}
    \right].
\end{equation}
The reported uncertainty was taken to be the sample standard deviation of
the resulting bootstrap distribution.


In Fig.~\ref{fig:fig2}(a), all six one-dimensional system sizes,
$L=64,96,144,216,324,$ and $486$, were included in the fits $
-\ln\langle X\rangle=a\ln L+c
$.  In Fig.~\ref{fig:fig2}(b), all ten two-dimensional sizes
$L=8,12,\ldots,44$ were retained. The ordinary-ensemble data were fitted to
$-\ln\langle X\rangle=aL+c$, whereas the two-replica data were fitted to
$-\ln\langle X\rangle=aL+b\ln L+c$. The area-law-subtracted data in the inset were fitted as linear functions of $\ln L$ over the same full size range.

All seven system sizes $L=64,128,192,256,384,512,$ and $768$ are displayed in Fig.~\ref{fig:fig4}, while the smallest sizes were excluded from the asymptotic fits. The charge variance was fitted to
$
\operatorname{Var}(q)=a_{\mathrm{var}}\ln L+b_{\mathrm{var}}
$
using $L\geq128$, and the number entropy was fitted to
$
S^{\mathrm{num}}
=
a_{\mathrm{num}}\ln(\ln L)+b_{\mathrm{num}}
$
over the same range. For the symmetry-resolved entropy, we used
$
S_2(q)-S_A^{(2)}
+\frac{1}{2}\ln(\ln L)
=
c_q+\frac{a_q}{\ln L}.
$
The $q=0$ fit used $L\geq128$, whereas the $q=1$ fit used $L\geq192$. 

The finite-size extrapolations of the reconstructed quantities in Fig.~\ref{fig:fig4} were performed using weighted least-squares fits, with the reconstructed Monte Carlo standard errors treated as absolute uncertainties. The quoted one-standard-deviation uncertainties of the fit parameters were obtained from the square roots of the diagonal elements of the resulting parameter covariance matrices and were not rescaled by the reduced chi-squared.

Evaluating all $M=L/2$ counting fields introduces an $\mathcal{O}(L)$ measurement overhead, which is subleading compared with the $\mathcal{O}(\beta L)$ cost of a standard SSE sweep in one dimension. Since we fix $\beta\propto L$, the leading computational cost per sweep scales as $\mathcal{O}(L^2)$. The two-replica calculation has the same asymptotic complexity, with an approximately twofold increase in computational work and memory because the imaginary-time configuration is replicated twice.

\end{document}